%% file: wpaper.tex
\documentclass[amssymb,twocolumn,aps,floatfix,showpacs,superscriptaddress]{revtex4-1}
\usepackage{graphicx}
\usepackage{bm}

\newcommand{\etal}{\emph{et al.}}
\usepackage{booktabs}
\usepackage{dcolumn}

\newcolumntype{d}[5]{D{.}{\cdot}{#1}}
\usepackage{multirow}
\usepackage{epstopdf}
\usepackage[version=4]{mhchem}
\usepackage{enumerate}
\usepackage{hyperref}
\hypersetup{
    colorlinks=true,    
    citecolor=blue,   
    linkcolor=blue,             
    urlcolor=blue          
}
\begin{document}
\title{Dirac R-matrix calculations for the electron-impact excitation of neutral tungsten providing noninvasive diagnostics for magnetic confinement fusion}
\author{R. T. Smyth}
\email{rsmyth41@qub.ac.uk}
\affiliation{School of Mathematics and Physics, Queen's University Belfast, Belfast BT7 1NN, Northern Ireland, UK}
\author{C. P. Ballance}
\affiliation{School of Mathematics and Physics, Queen's University Belfast, Belfast BT7 1NN, Northern Ireland, UK}
\author{C. A. Ramsbottom}
\affiliation{School of Mathematics and Physics, Queen's University Belfast, Belfast BT7 1NN, Northern Ireland, UK}
\author{C. A. Johnson}
\affiliation{Department of Physics, Auburn University, Auburn, Alabama 36849, USA}
\author{D. A. Ennis}
\affiliation{Department of Physics, Auburn University, Auburn, Alabama 36849, USA}
\author{S. D. Loch}
\affiliation{Department of Physics, Auburn University, Auburn, Alabama 36849, USA}

\begin{abstract}
Neutral tungsten is the primary candidate as a wall material in the divertor region of the International Thermonuclear Experimental Reactor (ITER). The efficient operation of ITER depends heavily on precise atomic physics calculations for the determination of reliable erosion diagnostics, helping to characterise the influx of tungsten impurities into the core plasma. The following paper presents detailed calculations of the atomic structure of neutral tungsten using the multiconfigurational Dirac-Fock method, drawing comparisons with experimental measurements where available, and includes a critical assessment of existing atomic structure data. We investigate the electron-impact excitation of neutral tungsten using the Dirac R-matrix method and, by employing collisional-radiative models, we benchmark our results with recent Compact Toroidal Hybrid measurements. The resulting comparisons highlight alternative diagnostic lines to the widely used 400.88nm line. 
\end{abstract}
\maketitle

\section{Introduction}\label{section-introduction}
The development and accuracy of fundamental atomic physics models for neutral tungsten (W I, $Z$=74) will be crucial to understand the needs of ITER. These theoretical models will grant a predictive capability for providing new temperature and density diagnostics for the ITER plasma. At present, existing models are either incomplete or of insufficient accuracy for diagnostic work. As a result, disagreements with experimental measurements are seen for a number of diagnostic lines in the 400-523 nm range \cite{ref1.5, ref1.23, ref1.6, ref1.7, ref1.18, ref1.22, ref1.4}. Difficulties with neutral tungsten diagnostics arise due to the ambiguity of line identification and notable absence of classification for many levels in the tungsten spectrum (given in many databases, including NIST \cite{ref1.24}), hindering the identification of spectral lines. Existing atomic structure models must also predict energies close to measured values to provide accurate wavelengths for spectral comparison. Large scale atomic structure models of Quinet \etal, \cite{ref1.25, ref1.26} have employed Cowan's code using elaborate configuration sets for this reason. However, plane-wave Born calculations \cite{ref1.21}, based upon a subset of these configurations, may be incomplete in terms of missing transitions. Additionally, existing semi-empirical calculations of Beigman \etal, \cite{ref1.4} (employing the van Regemorter formula \cite{ref3.5}) are also incomplete, considering only dipole transitions. As a result, neither would be expected to provide data of high accuracy, yet they represent the current models used for diagnostic purposes. Furthermore, due to the large number of spectral lines emitted, the issue of potential line blending must also be considered in the choice of lines used as diagnostics. For example, the predominantly used diagnostic line for neutral tungsten at 400.88 nm may be blended with a neighbouring W II line causing potential problems with observation. However, the work of Ekberg \etal, \cite{ref1.29} has demonstrated the separation of different ion stages of tungsten, indicating that the W II line is likely to be weak compared to the neutral tungsten line. Although this may be the case, the availability of alternative diagnostic lines would be highly beneficial but can only be provided through detailed and extensive atomic structure and collisional calculations.

In light of the issues mentioned, this work will address the disagreement in the literature with regards to the classification of fine structure levels. There has been little consensus in assigning a configuration to the upper $\ce{^7P^o_4}$ level of the important 400.88 nm line to which several authors have assigned the $\ce{{5d}^5{6p}}$ configuration \cite{ref1.9, ref1.10}. However, isotope shift measurements of Gluck \cite{ref1.19} and more recent measurements by Lee \etal, \cite{ref1.16} suggest that this upper level belongs to the $\ce{{5d}^4{6s}{6p}}$ configuration, as initially reported by \cite{ref1.11}. Recent {\it ab initio} calculations \cite{ref1.12} also suggest that the dominant configuration for this upper level is $\ce{{5d}^4{6s}{6p}}$. The present work will address this issue through large scale atomic structure calculations, and comparisons with recent Compact Toroidal Hybrid (CTH) \cite{ref1.8} measurements, in terms of spectral height and position, will provide independent validation of both theory and measurement.

In terms of its application, tungsten is the leading material choice for plasma facing components (PFCs) in the divertor region of ITER, with initial experimental investigations underway at JET to assess its suitability \cite{ref1.20, ref1.17, ref1.1}. However, PFCs in the divertor region will unavoidably come into contact with the plasma, resulting in an influx of tungsten impurities from the PFCs into the core plasma. Considering that the spectrum of neutral tungsten, a complex open-d shell system, is very dense, it becomes an efficient radiator at high temperatures. Small amounts of these impurities can contaminate the plasma and may ultimately result in a quenching of the thermonuclear fusion process. Thus, it is required that the concentration of such impurities be kept less than $10^{-5}$ \cite{ref1.2}. This impurity influx can be diagnosed through the use of the commonly used SXB spectral diagnostic \cite{ref1.3}, which is dependent upon accurate electron-impact excitation and electron-impact ionisation calculations for neutral tungsten. In addition, for an analysis of the effects metastable states on the influx diagnostics, a complete model which includes transitions between all metastables is required \cite{ref1.28}. However, as mentioned previously, existing models are incomplete and omit such transitions. In this paper we focus on the electron-impact excitation of neutral tungsten, providing a complete non-perturbative model to replace the existing plane-wave Born \cite{ref1.4} and semi-empirical \cite{ref1.21} calculations.

The remainder of the paper is structured as follows. In Section \ref{section-atomicstructure} we investigate the atomic structure and issues of level identification using the multiconfigurational Dirac-Fock method, implemented by GRASP$^0$ \cite{ref1.13, ref1.14}. In Section \ref{section-eie} we discuss our non-perturbative electron-impact excitation calculations using the Dirac atomic R-matrix codes (DARC) \cite{ref1.15}, the capabilities of which have been extended to handle the present calculations. Finally, in Section \ref{pops} we present our population modelling calculations. Results are benchmarked against recent CTH measurements and we offer alternative diagnostic lines to the widely used 400.88 nm line.

\section{Atomic Structure}\label{section-atomicstructure}
\begin{figure}
\includegraphics[width=0.4\textwidth]{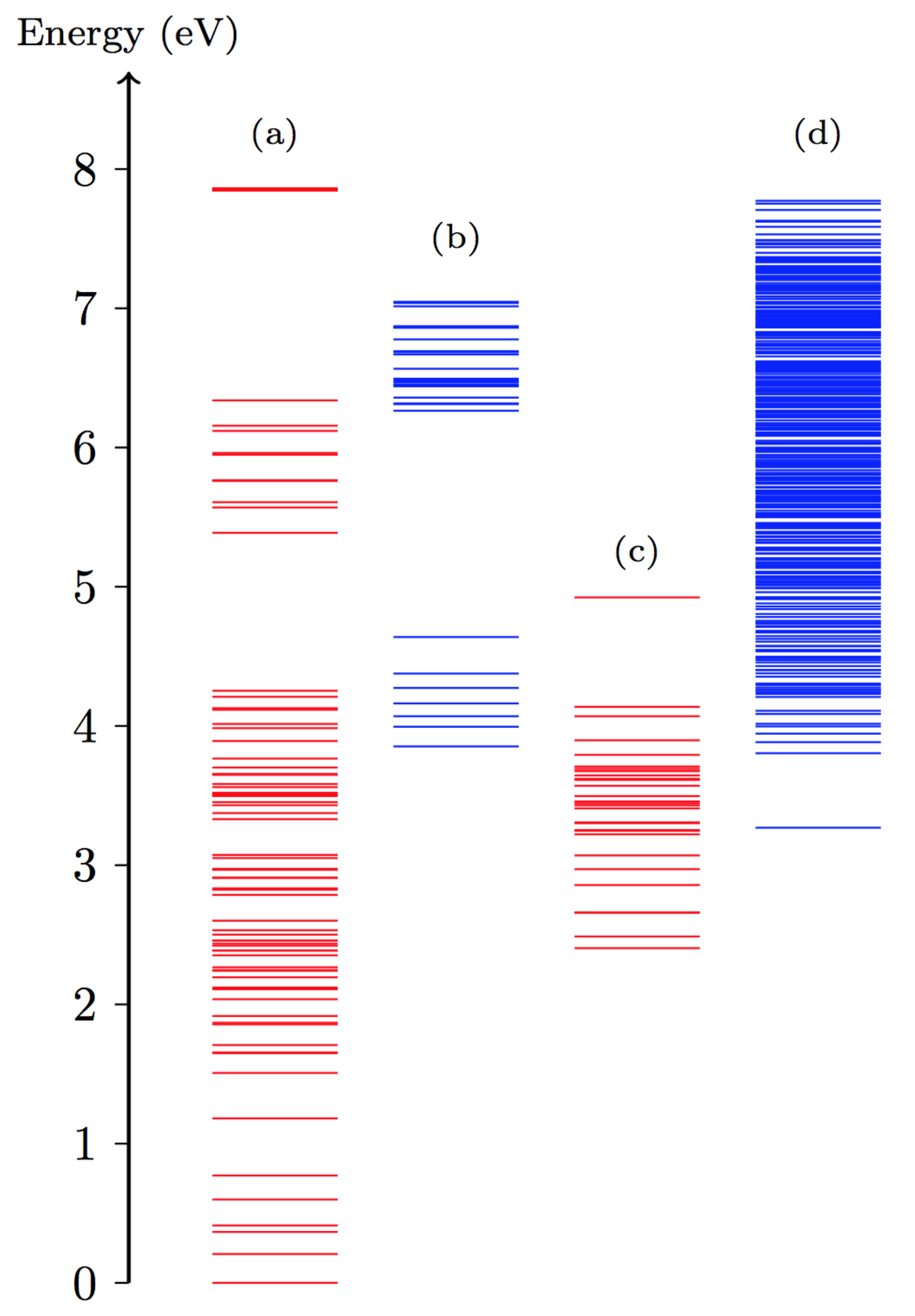}     
\caption{Energy level spectrum of neutral tungsten with each horizontal line representing an observed fine structure level. (a) Even parity levels with a configuration and term classification; (b) unclassified even levels; (c) classified odd levels; (d) unclassified odd levels. \label{energyspectrum}}
\end{figure}

The atomic structure of W I is highly complex, with half-open d subshells not only in its ground state configuration of \ce{{5d}^4{6s}^2}, but also in many of its excited state configurations, giving a very rich and complicated fine structure spectrum. This renders the theoretical modelling of neutral tungsten a formidable task. Further difficulties arise due to the lack of configuration and term labels given to the majority of observed fine structure levels, illustrated in Figure \ref{energyspectrum}, making comparisons between theory and experiment problematic beyond the first few eV of the atomic structure. We note that our goal is not an exhaustively converged calculation for all transitions, but rather completeness in terms of the strong transitions across the 200-500nm window.

Existing atomic structure models for W I have been determined using Cowan's relativistic Hartree-Fock (HFR) method \cite{ref1.27}. Calculations of Quinet \etal, \cite{ref1.25, ref1.26} have employed the HFR method using elaborate configuration expansions, including the effects of core-polarisation, and parametrically fitting calculated energies to the experimental values compiled by Kramida and Shirai \cite{ref2.1}. Similarly, calculations of Wyart \cite{ref1.12} have also employed the HFR method, although using a much smaller set of configurations, to supplement the parametric Racah-Slater approach. It is noted that these calculations have displayed fair general agreement with experimental values.

For our calculation we used the relativistic multi-configurational Dirac-Fock method, implemented by GRASP$^0$. We predominantly made use of the extended average level (EAL) method in which we give the diagonal elements of our Dirac-Coulomb Hamiltonian, defined (in atomic units) as
\begin{equation} \label{graspdirachamiltonian}
H_D = \sum_{i} \left( -ic {\boldsymbol{\alpha}} \cdot {\boldsymbol{\nabla}}_i +  (\beta-I_4)c^2 - \frac{Z}{r_i} \right) + \sum_{i > j} \frac{1}{r_{ij}},
\end{equation}
weights proportional to $(2J+1)$. A variational procedure then optimises the trace of the weighted Hamiltonian, and in turn, allows one to determine a set of atomic orbitals which describe closely lying states with good accuracy. In Eq.~(\ref{graspdirachamiltonian}) ${\boldsymbol{\alpha}}$ and $\beta$ are related to the set of Pauli spin matrices, $I_4$ is the $4 \times 4$ identity matrix, $Z$ is the atomic number, $c$ is the speed of light, $r_i$ denotes the position of electron $i$ and $r_{ij}=|{\bf r}_i -{\bf r}_j|$ is the inter-electron distance.

Careful consideration was given to the calculation and optimisation of our W I atomic structure. However, for complex neutral systems, GRASP$^0$ often has difficulty converging shells with a high principal quantum number. Therefore, it is common practice to first generate orbitals for a more highly ionised case further up the isoelectronic sequence. These orbitals then form the initial guess to start the Dirac-Hartree-Fock iterative process for the neutral case. We have adopted this methodology within the present calculations using orbitals from Re II ($Z=75$). 

To determine our atomic structure for W I we allowed all orbitals to be variationally determined by invoking the EAL method (first for Re II, then for W I as discussed above) with the following 21 configurations:
$\ce{{5d}^4} \{ \ce{{6s}^2}, \ce{{6p}^2}, \ce{{6d}^2}, \ce{{6s}{6p}}, \ce{{6s}{6d}}, \ce{{7s}^2} \}$;
$\ce{{5d}^5} \{ \ce{{6s}}, \ce{{6p}}, \ce{{6d}}, \ce{{7s}} \}$;
$\ce{{5d}^3} \{ \ce{{6s}^2{6d}}, \ce{{6d}^3}, \ce{{6s}^2{7s}} \}$;
$\ce{{5d}^6}$;
$\ce{{5d}{6s}^2{6d}^3}$;
$\ce{{5p}^5} \{ \ce{{5d}^7}, \ce{{5d}^6{6s}} \}$;
$\ce{{5p}^4} \{ \ce{{5d}^8}, \ce{{5d}^6{6s}^2}, \ce{{5d}^7{6s}} \}$;
and finally the $\ce{{5s}{5p}^6{5d}^7}$ configuration. We then hold all existing orbitals fixed and introduce and optimise on a $\ce{{7p}}$ orbital. The EAL method was employed with the $\ce{{5d}{6s}^2{6d}^3}$ and $\ce{{5d}^3{6s}^2{6d}}$ configurations removed and the $\ce{{5d}^4{7p}^2}$ and $\ce{{5d}^5{7p}}$ configurations included. 
Next, all restrictions on our orbitals were removed, the \ce{{5d}^4{6s}{7s}} configuration was included, and a further EAL calculation was carried out.
We then optimised the $\ce{{6d}}$, $\ce{{7s}}$, and $\ce{{7p}}$ orbitals by holding the remaining orbitals fixed and including the $\ce{{5d}^3{6s}}^2 \{ \ce{{6p}}, \ce{{6d}}, \ce{{7p}} \}$ configurations for one last EAL calculation.
Finally, holding all orbitals fixed and carrying out a single configuration interaction (CI) calculation yields the W I atomic structure taken through to the Dirac R-matrix scattering calculation. 

\begin{table}
\begin{tabular}{ l l c c c }
\hline
No & Level & Expt \cite{ref2.1} & GRASP$^0$ &$|\Delta E |$ \\  
\hline
1 & \ce{{5d}^4{6s}^2 ^5D_0}& 0.00000 & 0.00000 & 0.00000\\ 
2 &\ce{{5d}^4{6s}^2 ^5D_1}& 0.01522 & 0.00936 & 0.00586\\
3 &\ce{{5d}^5(^6S){6s} ^7S_3}& 0.02689 & 0.02983 & 0.00294\\ 
4 &\ce{{5d}^4{6s}^2 ^5D_2}& 0.03030 & 0.02152 & 0.00878\\
5 &\ce{{5d}^4{6s}^2 ^5D_3}& 0.04401 & 0.03477 & 0.00924 \\ 
6 &\ce{{5d}^4{6s}^2 ^5D_4}& 0.05667 & 0.04876 & 0.00791\\
7 &\ce{{5d}^4{6s}^2 ^3P_0}& 0.08683 & 0.10237 & 0.01555 \\ 
8 &\ce{{5d}^4{6s}^2 ^3H_4}& 0.11083 & 0.13697 & 0.02614\\
9 &\ce{{5d}^4{6s}^2 ^3P_1}& 0.12126 & 0.13705 & 0.01579\\ 
10 &\ce{{5d}^4{6s}^2 ^3G_3}& 0.12164 & 0.14721 & 0.02577\\
11 &\ce{{5d}^4{6s}^2 ^3F_2}& 0.12555 & 0.15237 & 0.02682\\ 
12 &\ce{{5d}^4{6s}^2 ^3D_2}& 0.13647 & 0.16166 & 0.02519\\ 
13 &\ce{{5d}^4{6s}^2 ^3H_5}& 0.13733 & 0.16179 & 0.02446\\ 
14 &\ce{{5d}^4{6s}^2 ^3D_3}& 0.14088 & 0.17371 & 0.03283\\
15 &\ce{{5d}^4{6s}^2 ^3G_4}& 0.14973 & 0.17926 & 0.02953\\ 
16 &\ce{{5d}^4{6s}^2 ^3H_6}& 0.15499 & 0.17959 & 0.02460\\
17 &\ce{{5d}^4{6s}^2 ^3F_4}& 0.15589 & 0.17179 &  0.01590\\
18 &\ce{{5d}^4{6s}^2 ^3F_3}& 0.16131 & 0.18253 & 0.02122\\ 
 & & & & \\
19 &\ce{{5d}^4{6s}(^6D){6p} ^7F^o_0}& 0.17669 & 0.15519 & 0.02150\\ 
20 &\ce{{5d}^4{6s}(^6D){6p} ^7F^o_1}& 0.18284 & 0.16060 & 0.02224\\
21 &\ce{{5d}^4{6s}(^6D){6p} ^7F^o_2}& 0.19546 & 0.17048 & 0.02498\\ 
22 &\ce{{5d}^4{6s}(^6D){6p} ^7D^o_1}& 0.19550 & 0.17862 & 0.01689\\
23 &\ce{{5d}^4{6s}(^6D){6p} ^7F^o_3}& 0.21002 & 0.18382 & 0.02620\\ 
24 &\ce{{5d}^4{6s}(^6D){6p} ^7D^o_2}& 0.21838 & 0.19354 & 0.02484\\
25 &\ce{{5d}^4{6s}(^6D){6p} ^7F^o_4}& 0.22566 & 0.20021 & 0.02545\\ 
26 &\ce{{5d}^3{6s}^2(^4F){6p} ^5F^o_1}& 0.23678 & 0.23610 & 0.00068\\
27 &\ce{{5d}^4{6s}(^6D){6p} ^7D^o_3}& 0.23865 & 0.20752 & 0.03113\\ 
28 &\ce{{5d}^4{6s}(^6D){6p} ^7P^o_2}& 0.23902 & 0.21836 & 0.02067\\
29 &\ce{{5d}^3{6s}^2(^4F){6p} ^5G^o_2}& 0.24028 & 0.24431 & 0.00404 \\ 
30 &\ce{{5d}^4{6s}(^6D){6p} ^5D^o_0}& 0.24267 & 0.25311 & 0.01044 \\
31 &\ce{{5d}^4{6s}(^6D){6p} ^7F^o_5}& 0.24309 & 0.22062 & 0.02248 \\
32 &\ce{{5d}^4{6s}(^6D){6p} ^7P^o_3}& 0.25049 & 0.22902 & 0.02147\\  
33 &\ce{{5d}^3{6s}^2(^4F){6p} ^5F^o_2}& 0.25208 & 0.24724 & 0.00484 \\ 
34 &\ce{{5d}^4{6s}(^6D){6p} ^5D^o_1}& 0.25314 & 0.26000 & 0.00686\\
35 &\ce{{5d}^4{6s}(^6D){6p} ^7P^o_4}& 0.25415 & 0.21907 & 0.03508\\ 
36 &\ce{{5d}^4{6s}(^6D){6p} ^5P^o_1}& 0.25697 & 0.24266 & 0.01431\\
37 &\ce{{5d}^4{6s}(^6D){6p} ^5P^o_2}& 0.26785 & 0.25654 & 0.01131\\
38 &\ce{{5d}^4{6s}(^6D){6p} ^5P^o_3}& 0.27873 & 0.27411 & 0.00462\\
39 &\ce{{5d}^4{6s}(^6D){6p} ^5F^o_4}& 0.28644 & 0.27603 & 0.01041\\
 & & & & \\
40 &\ce{{5d}^4{6s}{6p} ^3F^o_2}& 0.32561 & 0.34535 & 0.01974\\ 
41 &\ce{{5d}^5{6p} ^7P^o_3}& 0.33602 & 0.33973 & 0.00371\\ 
42 &\ce{{5d}^5{6p} ^7P^o_2}& 0.33630 & 0.33511 & 0.00118\\ 
43 &\ce{{5d}^5{6p} ^5P^o_1}& 0.35706 & 0.37371 & 0.01665\\ 
44 &\ce{{5d}^5{6p} ^7P^o_4}& 0.36982 & 0.35549 & 0.01433\\ 
45 &\ce{{5d}^4{6s}{6p} ^5F^o_4}& 0.41799 & 0.43109 & 0.01310\\ 
\hline
\end{tabular}
\caption{Fine structure energies of W I, in Rydbergs, obtained from the GRASP$^0$ model (relative to the ground state) compared to the experimental values compiled by Kramida and Shirai, \cite{ref2.1}. Absolute energy differences are given in the final column. \label{wenergytable1}} 
\end{table}

\begin{table}
\begin{tabular}{c c c c}
\hline
& Transition & \multicolumn{2}{ c } {$A_{ji}$ ($s^{-1}$)}\\
\cmidrule{3-4}
$\lambda$ (nm) & $j-i$ & Current & Expt \cite{ref2.1}\\   
\hline
243.60 &  45 - 5  & $2.80 \times 10^8$ & $2.54 \times 10^8$  \\
255.13 &  43 - 1  & $1.41 \times 10^8$ & $1.78 \times 10^8$  \\
265.65 &  44 - 3  & $2.51 \times 10^8$ & $6.70 \times 10^7$  \\ 
293.50 &  40 - 2  & $4.51 \times 10^7$ & $1.50 \times 10^7$  \\
294.44 &  42 - 3  & $1.52 \times 10^8$ & $1.08 \times 10^8$  \\
294.70 &  41 - 3  & $7.01 \times 10^7$ & $8.20 \times 10^7$  \\
375.79 &  39 - 5  & $2.09 \times 10^7$ & -  \\
376.84 &  36 - 2  & $2.20 \times 10^7$ & $3.47 \times 10^6$  \\
383.51 &  37 - 4  & $2.63 \times 10^7$ & $5.20 \times 10^6$  \\
384.62 &  33 - 2  & $1.46 \times 10^7$ & $2.14 \times 10^6$  \\
400.88 &  35 - 3  & $1.14 \times 10^7$ & $1.63 \times 10^7$  \\
407.44 &  32 - 3  & $5.66 \times 10^6$ & $1.00 \times 10^7$  \\
410.27 &  38 - 6 & $5.39 \times 10^6$  & $4.90 \times 10^6$ \\
429.46 &  28 - 3  & $3.17 \times 10^7$ & $1.24 \times 10^7$  \\
430.21 &  27 - 3 & $4.54 \times 10^6$  & $3.60 \times 10^6$ \\
488.69 &  31 - 6  & $1.48 \times 10^6$ & $8.10 \times 10^5$  \\ 
498.26 &  20 - 1 &  $2.13 \times 10^5$ & $4.17 \times 10^5$  \\ 
505.33 &  22 - 2  & $3.24 \times 10^6$ & $1.90 \times 10^6$  \\ 
522.47 &  24 - 5  & $1.23 \times 10^6$ & $1.20 \times 10^6$  \\ 
\hline
\end{tabular} 
\caption{Radiative transition rates (in $s^{-1}$) obtained from the present model compared to experimental values compiled by Kramida and Shirai, \cite{ref2.1}. The rates shown are obtained after shifting energies to experimental values. 
\label{wavaluetable1}}
\end{table}

This detailed calculation resulted in a large 25 configuration, 7825 level atomic structure model. Retaining the first 250 levels in the close-coupling expansion of our wavefunction in the Dirac R-matrix calculation was sufficient to encapsulate the strong transitions across the 200-500nm window. A sample of fine structure levels from the atomic structure calculation are shown in Table \ref{wenergytable1} to give an indication of the quality and accuracy of our model. A selection of highly excited levels involved in transitions we believe to be reliable for future diagnostic work are also given (numbered 40-45 in Table \ref{wenergytable1}). In Table \ref{wavaluetable1} we present some of the strongest transitions amongst levels given in Table \ref{wenergytable1} which will be useful for future diagnostic purposes. Results are compared to the experimental values compiled by Kramida and Shirai, \cite{ref2.1} and the rates shown have been obtained after energies were shifted to experimental thresholds. This energy level shifting is carried through to our scattering calculation and ensures that calculated wavelengths will agree with experimental thresholds.

The large amount of CI, paired with the absence of many level classifications, made comparisons between theory and experiment very difficult. Furthermore, in many cases the mixing was so severe that term assignments have little physical meaning. Thus, a better description of these states would be provided by looking at the compositions of their eigenvectors. For the majority of odd parity levels, little consensus of classification has been achieved due to the strong mixing between the \ce{{5d}^5{6p}}, $\ce{{5d}^4{6s}{6p}}$, and \ce{{5d}^3{6s}^2{6p}} configurations. We address this issue with the support of recent {\it ab initio} calculations of Wyart \cite{ref1.12} and is best illustrated by considering the sample of levels given in Table \ref{wenergytable1}. Referring to Table \ref{wenergytable1} we have assigned the \ce{{5d}^3{6s}^2{6p}} configuration to the \ce{^5F^o_{1,2}} odd levels (numbered 26 and 33 in Table \ref{wenergytable1}) as opposed to the $\ce{{5d}^4{6s}{6p}}$ configuration assignment given previously by \cite{ref1.10}. The \ce{^5F^o_{1}} assignment is supported by recent calculations of \cite{ref1.12}. However, our \ce{^5F^o_{2}} assignment disagrees with the calculations of \cite{ref1.12}, who finds that the $\ce{{5d}^4{6s}{6p}}$ is dominant over the \ce{{5d}^3{6s}^2{6p}} configuration by a small fraction.
We also note that an additional level (numbered 29 in Table \ref{wenergytable1}) has been assigned the \ce{{5d}^3{6s}^2{6p}} configuration and given a new term label, supported by the work of \cite{ref1.12}. 
Finally, we have assigned the $\ce{^7P^o_{2,3,4}}$ levels (numbered 28, 32, and 35 in Table \ref{wenergytable1}) the $\ce{{5d}^4{6s}{6p}}$ configuration, as opposed to the \ce{{5d}^5{6p}} configuration given previously by \cite{ref1.9} and \cite{ref1.10}, and again this is supported by the work of \cite{ref1.11} and \cite{ref1.12}. 

For such a highly complex neutral atomic system, we see very good overall agreement with experiment. Compared with experimental values, our fine structure levels have an average percentage error of approximately 11\% where the largest sources of error come from the even parity levels. These have errors ranging from 10\% up to 38\% with an overall average error of 18\%. However, the odd parity levels are very well represented, with percentage errors ranging from 0.02\% up to 14\% with an overall average error of 5\%.

Now referring to the radiative transition rates listed in Table \ref{wavaluetable1} we see very good agreement for the $\ce{{5d}^4{6s}{6p} ^5F^o_4} \rightarrow \ce{{5d}^4{6s}^2 ^5D_3} $ transition (numbered $45 \rightarrow 5$), which has a percentage difference of 9.7\% when compared with experimental observation. Furthermore, good agreement with experiment is seen for the $\ce{{5d}^5{6p} ^5P^o_1} \rightarrow \ce{{5d}^4{6s}^2 ^5D_0}  $ (numbered $43 \rightarrow 1$) transition, having a difference of 23.2\%. For the important $400.88$nm, $\ce{{5d}^4{6s}{6p} ^7P^o_4} \rightarrow \ce{{5d}^5{6s} ^7S_3}$, diagnostic line (numbered $35 \rightarrow 3$) we see reasonable agreement with experiment, with a percentage difference of 35.4\%. In addition, the $\ce{{5d}^4{6s}{6p} ^5P^o_3} \rightarrow \ce{{5d}^4{6s}^2 ^5D_4}$ transition (numbered $38 \rightarrow 6$) exhibits very good agreement with experiment having a percentage difference of 9.5\%. Finally, excellent agreement is seen for the $\ce{{5d}^4{6s}{6p} ^7D^o_2} \rightarrow \ce{{5d}^4{6s}^2 ^5D_3}$ transition (numbered $24 \rightarrow 5$) with a difference of 0.02\%.

This very good overall agreement with experimental observation gives confidence in the accuracy and precision of the present W I target model, which has been carried through to the Dirac R-matrix scattering calculation as detailed in the next section. 

\section{Electron-impact Excitation}\label{section-eie}
\subsection{R-Matrix Theory}\label{section-theory}
First we note that a more detailed account of Dirac R-matrix theory is given elsewhere in \cite{ref3.1} and here we simply give an overview of the most salient facts. We begin with the division of configuration space into two distinct regions separated by an R-matrix boundary at $r=a$, chosen to entirely encapsulate the charge distribution of the $N$-electron target. Within the internal region, electron exchange and short range correlation effects between the incident electron and the target must be taken into account. Here, we expand the wavefunction as
\begin{equation}
\Psi_{jE}^{\Gamma}({\bf X}_{N+1})=\sum_k  A_{jk}^{\Gamma}(E) \psi_k^{\Gamma}({\bf X}_{N+1}),
\end{equation}
where $A_{jk}^{\Gamma}(E)$ are energy dependent coefficients and the energy independent basis functions are written as a close-coupling expansion:
\begin{multline} \label{internalwavefunction}
\psi_k^{\Gamma} =\mathcal{A}  \sum_i \sum_j \bar{\Phi}_i ^{\Gamma}({\bf X}_N; \hat{{\bf r}}_{N+1}\sigma_{N+1}) r_{N+1}^{-1} u_{ij}(r_{N+1}) a_{ijk}^{\Gamma} \\  + \sum_i \chi_i^{\Gamma}({\bf X}_{N+1}) b_{ik}^{\Gamma}. ~~~~~~~~~~~~~~~~~~~~~~~
\end{multline}
Here, $\Gamma =J\pi$ are the conserved quantum numbers, ${\bf X}_{N+1}={\bf x}_1, ..., {\bf x}_{N+1}$ (where ${\bf x}_i ={\bf r}_i \sigma_i$) are the set of space and spin coordinates of the $N+1$ electrons, $\bar{\Phi}_i^{\Gamma}$ are channel functions, $u_{ij}$ are functions which describe the continuum of the $(N+1)$ electron system for each value of angular momentum $J$, square integrable functions $\chi_i^{\Gamma}$ describe short range correlation effects, and the coefficients $a_{ijk}^{\Gamma}$ and $b_{ik}^{\Gamma}$ are determined from the diagonalisation of the $(N+1)$ electron Hamiltonian [analogous to Equation (\ref{graspdirachamiltonian})] over the energy independent basis.

Within the external region, the electron moves only in the long range potential of the target and we may neglect electron exchange and correlation effects. The wavefunction here takes the form 
\begin{equation} \label{externalwavefunction}
\Psi_{jE}^{\Gamma}= \sum_i \bar{\Phi}_i ^{\Gamma}({\bf X}_N; \hat{{\bf r}}_{N+1}\sigma_{N+1}) r_{N+1}^{-1} F_{ij}^{\Gamma}(r_{N+1}),
\end{equation}
where $\bar{\Phi}_i^{\Gamma}$ are identical channel functions as those in Equation (\ref{internalwavefunction}) and $F_{ij}^{\Gamma}$ are reduced radial functions. Matching this to asymptotic boundary conditions given by Young and Norrington \cite{ref3.2} allows one to determine the collision strengths ($\Omega_{ij}$) for an excitation from some initial level $i$ to some final level $j$.

\subsection{Scattering Calculation and Results}\label{section-scatteringcalculation} 
As mentioned in Section \ref{section-atomicstructure}, 250 levels of our 25 configuration, 7825 level atomic structure were retained in the close-coupling expansion of our energy independent R-matrix basis, requiring an R-matrix boundary at 32.96 atomic units, producing up to 1698 coupled channels and requiring the diagonalisation of matrices of sizes up to $43186 \times 43168$. Due to the sheer size of the overall atomic structure it was required that integer*8 capabilities be implemented throughout the Dirac atomic R-matrix codes in order to handle calculations of up to an order of magnitude larger.

We noted in Section \ref{section-introduction} that currently available electron-impact excitation models consisted of either plane-wave Born (PWB) \cite{ref1.21} or semi-empirical \cite{ref1.4} calculations. These plane-wave Born calculations have been based upon a subset of configurations used for large scale HFR calculations of Quinet \etal, \cite{ref1.25, ref1.26} while the semi-empirical calculations are based upon the van Regemorter formula \cite{ref3.5}, using experimental data from the NIST database. However, datasets resulting from these calculations are incomplete, with PWB calculations omitting transitions between large numbers fine structure levels and the other considering only dipole transitions. To the best of our knowledge the present calculation represents the first complete, non-perturbative calculation for the electron-impact excitation of neutral tungsten.

In this work we have carried out calculations for 60 $J\pi$ partial waves from $2J=1$ up to and including $2J=59$. A continuum basis size of 20 was used for lower partial waves $2J=1$ to $2J=33$ and a smaller basis size of 15 was used for the remainder. This choice of basis was more than sufficient to span a large energy range of $0-30$ eV. A fine mesh of 5000 points was used for all partial waves $2J=1$ to $2J=25$ with an energy spacing of $4.38 \times 10^{-4}$ Ryd and a coarser mesh of 1000 points with an energy spacing of $2.19 \times 10^{-3}$  Ryd was used for all partial waves $2J=27$ up to $2J=59$. For higher partial waves $2J>59$ a ``top-up'' procedure described by Burgess \cite{ref3.3} was employed to estimate the collision strengths for all dipole transitions.

\begin{figure*}
\includegraphics[width=0.78\textwidth]{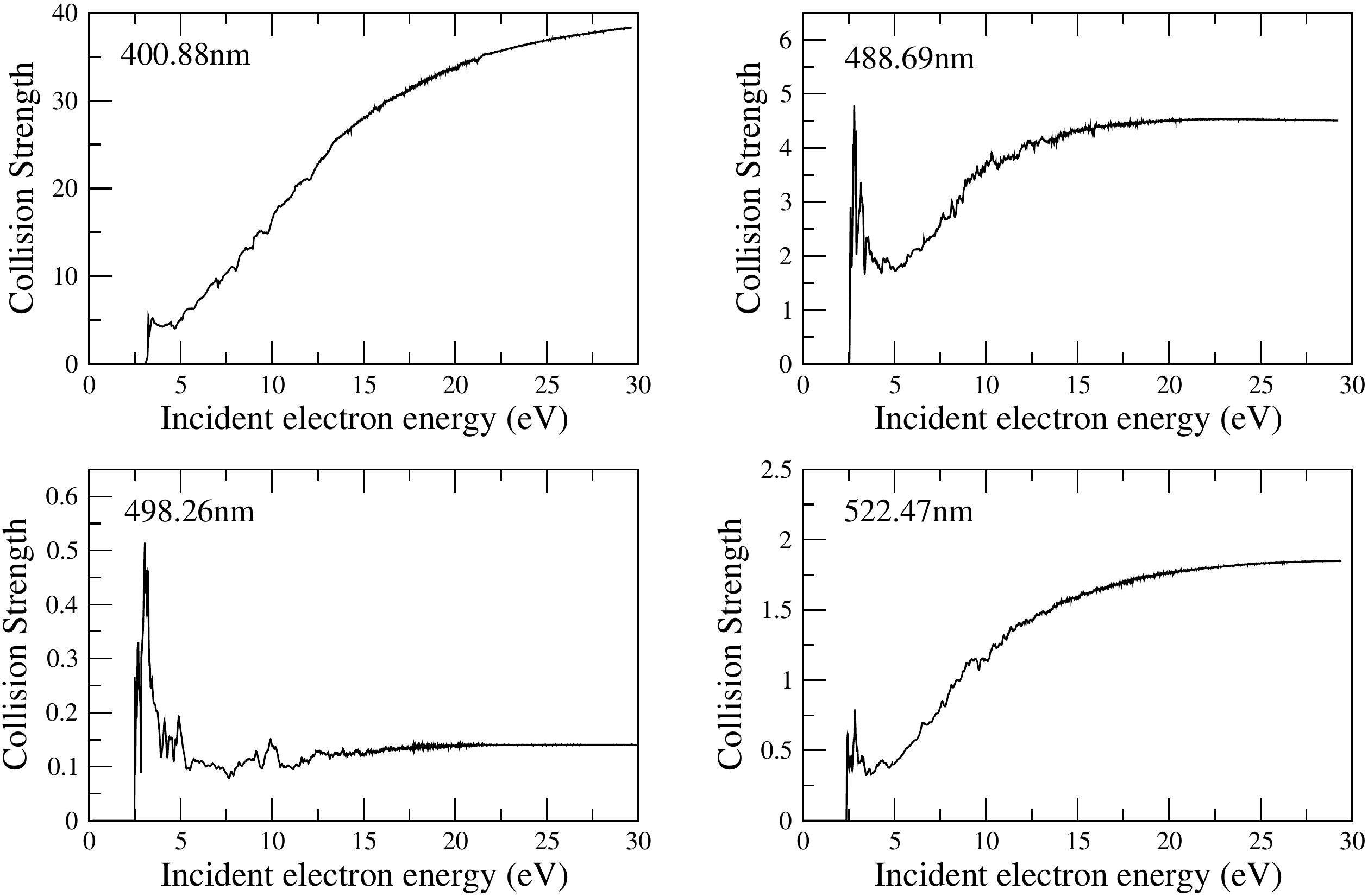}     
\caption{Plot showing collision strengths for the 
$\ce{{5d}^4{6s}{6p}} \ce{ ^7P^o_4} \rightarrow \ce{{5d}^5{6s}} \ce{ ^7S_3}$ (400.88nm), 
$\ce{{5d}^4{6s}{6p}} \ce{ ^7F^o_5} \rightarrow \ce{{5d}^4{6s}^2} \ce{ ^5D_4}$ (488.69nm), 
$\ce{{5d}^4{6s}{6p}} \ce{ ^7F^o_1} \rightarrow  \ce{{5d}^4{6s}^2} \ce{ ^5D_0}$ (498.26nm), and 
$\ce{{5d}^4{6s}{6p}} \ce{ ^7D^o_2} \rightarrow \ce{{5d}^4{6s}^2} \ce{ ^5D_3}$ (522.47nm)
transitions from the present R-matrix calculation. \label{collisionplot}}
\end{figure*}

In Fig \ref{collisionplot} we present a selection of results from the present R-matrix calculation. We specifically present collision strengths for the $\ce{{5d}^4{6s}{6p}} \ce{ ^7P^o_4} \rightarrow \ce{{5d}^5{6s}} \ce{ ^7S_3}$ (400.88nm), $\ce{{5d}^4{6s}{6p}} \ce{ ^7F^o_5} \rightarrow \ce{{5d}^4{6s}^2} \ce{ ^5D_4}$ (488.69nm), $\ce{{5d}^4{6s}{6p}} \ce{ ^7F^o_1} \rightarrow  \ce{{5d}^4{6s}^2} \ce{ ^5D_0}$ (498.26nm) and $\ce{{5d}^4{6s}{6p}} \ce{ ^7D^o_2} \rightarrow \ce{{5d}^4{6s}^2} \ce{ ^5D_3}$ (522.47nm) transitions which have been under investigation as potential diagnostic lines \cite{ref1.4,ref1.5, ref1.23, ref1.6, ref1.7, ref1.18, ref1.22}. Although very little scattering data for neutral tungsten is currently available in the literature to readily compare with the current collision strengths, existing datasets are available in the form of effective collision strengths obtained from PWB calculations (as discussed above). We recognise that scattering data obtained from semi-empirical calculations (again, as discussed above) also exists, but we will not consider these any further and restrict our discussion to PWB results. Thus, we use the present R-matrix collision strengths to calculate the effective collision strengths, defined as
\begin{equation}
\Upsilon_{ij}(T_e)=\int^{\infty}_{0} \Omega_{ij}\exp\left( {-\frac{\epsilon_j}{kT_e}} \right) d \left(\frac{\epsilon_j}{kT_e} \right),
\end{equation}
where $\epsilon_j$ is the energy of the scattered electron, $k$ is Boltzmann's constant and $T_e$ is the electron temperature in Kelvin.

\begin{figure*} 
\includegraphics[width=0.78\textwidth]{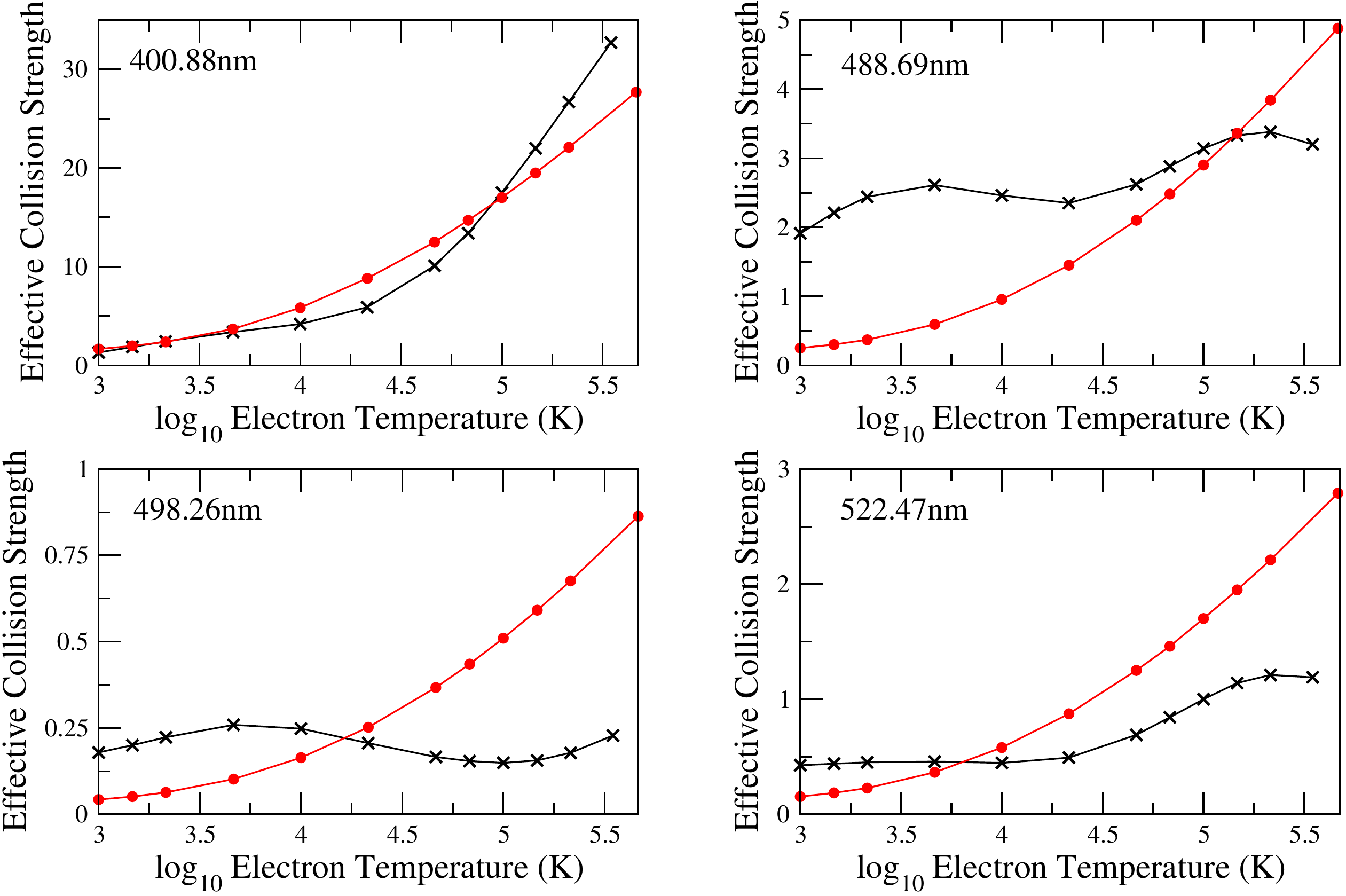}     
\caption{Plot showing effective collision strengths for the 
$\ce{{5d}^4{6s}{6p}} \ce{ ^7P^o_4} \rightarrow \ce{{5d}^5{6s}} \ce{ ^7S_3}$ (400.88nm), 
$\ce{{5d}^4{6s}{6p}} \ce{ ^7F^o_5} \rightarrow \ce{{5d}^4{6s}^2} \ce{ ^5D_4}$ (488.69nm), 
$\ce{{5d}^4{6s}{6p}} \ce{ ^7F^o_1} \rightarrow  \ce{{5d}^4{6s}^2} \ce{ ^5D_0}$ (498.26nm), and 
$\ce{{5d}^4{6s}{6p}} \ce{ ^7D^o_2} \rightarrow \ce{{5d}^4{6s}^2} \ce{ ^5D_3}$ (522.47nm)
transitions. The solid black line with crosses are the current R-matrix results and the red line with dots are results from a plane-wave Born calculation. \label{400500plot}}
\end{figure*}

We see from Fig \ref{400500plot} that in the case of the 400.88nm transition the R-matrix and PWB collision strengths diverge as the electron temperature increases. At low temperatures there is reasonable agreement but at higher temperatures the R-matrix collision strengths become $\sim$25\% larger than the PWB results in the temperature range shown. For the 522.47nm transition we see reasonable agreement at low temperatures. However, as the electron temperature increases the PWB and R-matrix results begin to diverge, with the PWB effective collision strength becoming up to a factor of $\sim$3 larger than the current R-matrix results. The most striking differences can be seen for the 488.69nm and 498.26nm transitions where it is clear there is very little agreement. For the 488.69nm transition, it is evident that there are large disagreements at low temperatures. Here, the present R-matrix results are up to a factor of $\sim$7 larger than the PWB effective collision strengths. As the electron temperature increases this disagreement lessens but it appears the results begin to diverge at higher temperatures. For the 498.26nm transition there is slight agreement at lower temperatures, but as the electron temperature increases the PWB effective collision strengths become up to $\sim$3.5 times larger than the current results. 

Overall it is clear that there are, in general, appreciable differences between the current R-matrix results and existing PWB results. For the spin-changing transitions (488.69nm, 498.26nm, and 522.47nm) these differences will arise due to the fact that the PWB method is poor at accurately calculating collision strengths for transitions of this type. In addition, for all transitions, differences will arise due to the variations in atomic structures employed in each scattering calculation. 

\subsection{Convergence and Accuracy}
Deducing the accuracy of the present scattering calculation is difficult given the lack of available non-perturbative data in the literature. However, here we analyse the effects of including successively higher $J\pi$ partial waves on the collision strengths and compare the results of our 250 state calculation with an additional sample 200 state calculation to clearly demonstrate convergence and accuracy of the results presented in the previous section. We use two representative examples for this: the strong 400.88nm, $\ce{{5d}^4{6s}{6p}} \ce{ ^7P^o_4} \rightarrow \ce{{5d}^5{6s}} \ce{ ^7S_3}$ dipole transition, and the 488.69nm, $\ce{{5d}^4{6s}{6p}} \ce{ ^7F^o_5} \rightarrow \ce{{5d}^4{6s}^2} \ce{ ^5D_4}$ spin-changing transition.

\begin{figure}[ht!]
\includegraphics[width=0.45\textwidth]{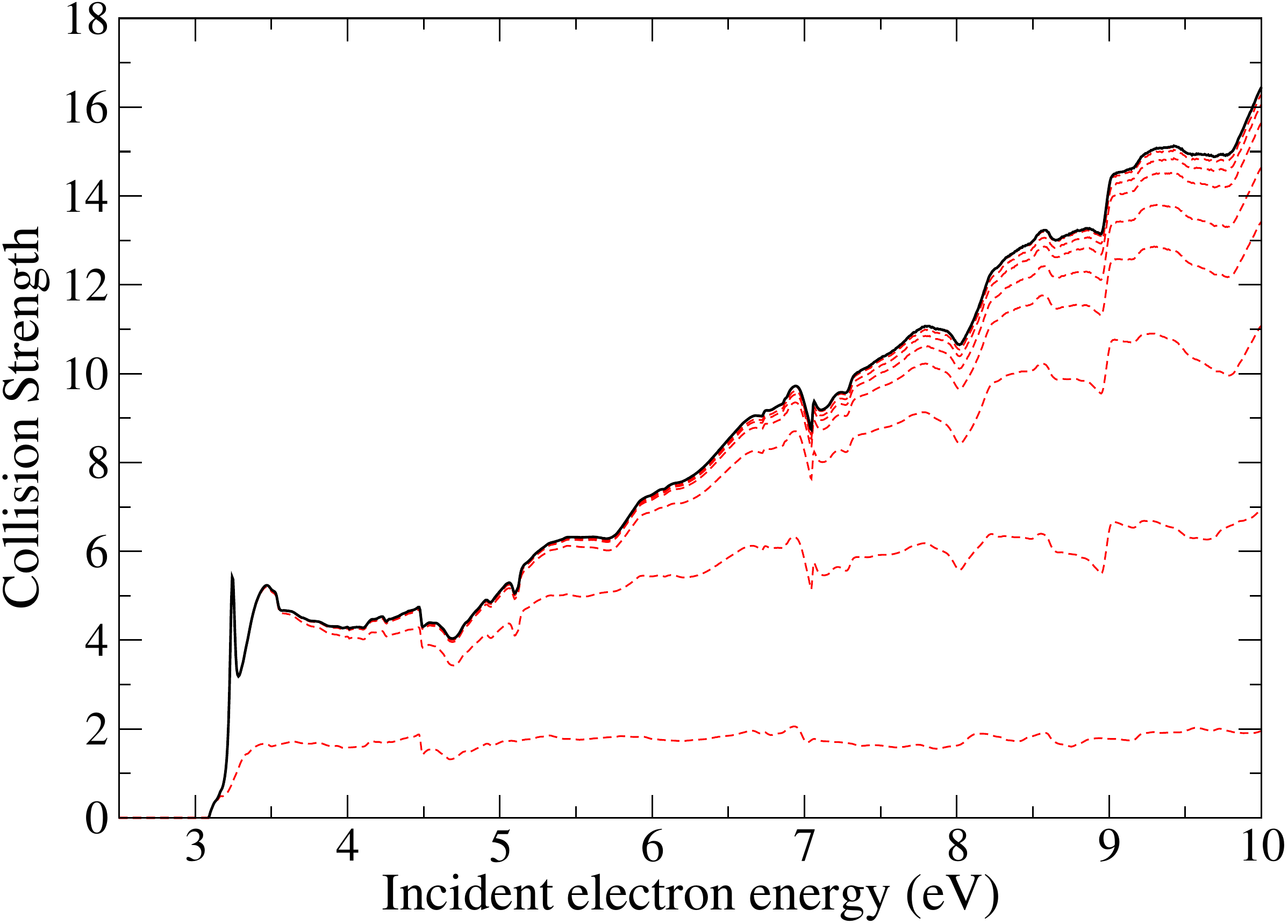}     
\caption{Plot showing successively higher partial wave contributions to the collision strength of the $\ce{{5d}^4{6s}{6p}} \ce{ ^7P^o_4} \rightarrow \ce{{5d}^5{6s}} \ce{ ^7S_3}$ transition. For each of the dashed red lines (going from bottom to top) we have included partial waves up to $2J=9, 13, 17, 21, 25, 33, 41, 47,$ and $53$ respectively. The solid black line is the current collision strength with all partial wave contributions included. \label{400.8convergence}}
\end{figure}

In Fig \ref{400.8convergence} we present the results of sample calculations showing the breakdown of how the inclusion of successively higher partial waves affects the collision strength of the $\ce{{5d}^4{6s}{6p}} \ce{ ^7P^o_4} \rightarrow \ce{{5d}^5{6s}} \ce{ ^7S_3}$ transition. As expected, the lowest $J\pi$ partial waves provide the largest contributions, which then lessen as the value of $J$ increases. Convergence can be best illustrated by considering the effect of the final six partial waves (such that $2J=55$, $57$, $59$), for which their inclusion results in a very small 0.2\% increase in the collision strength. Higher partial waves ($2J>59$) are expected to give even smaller contributions. This breakdown indicates that we have obtained near-converged collision strengths in terms of the partial waves included in our scattering calculation.

\begin{figure}[ht!]
\includegraphics[width=0.45\textwidth]{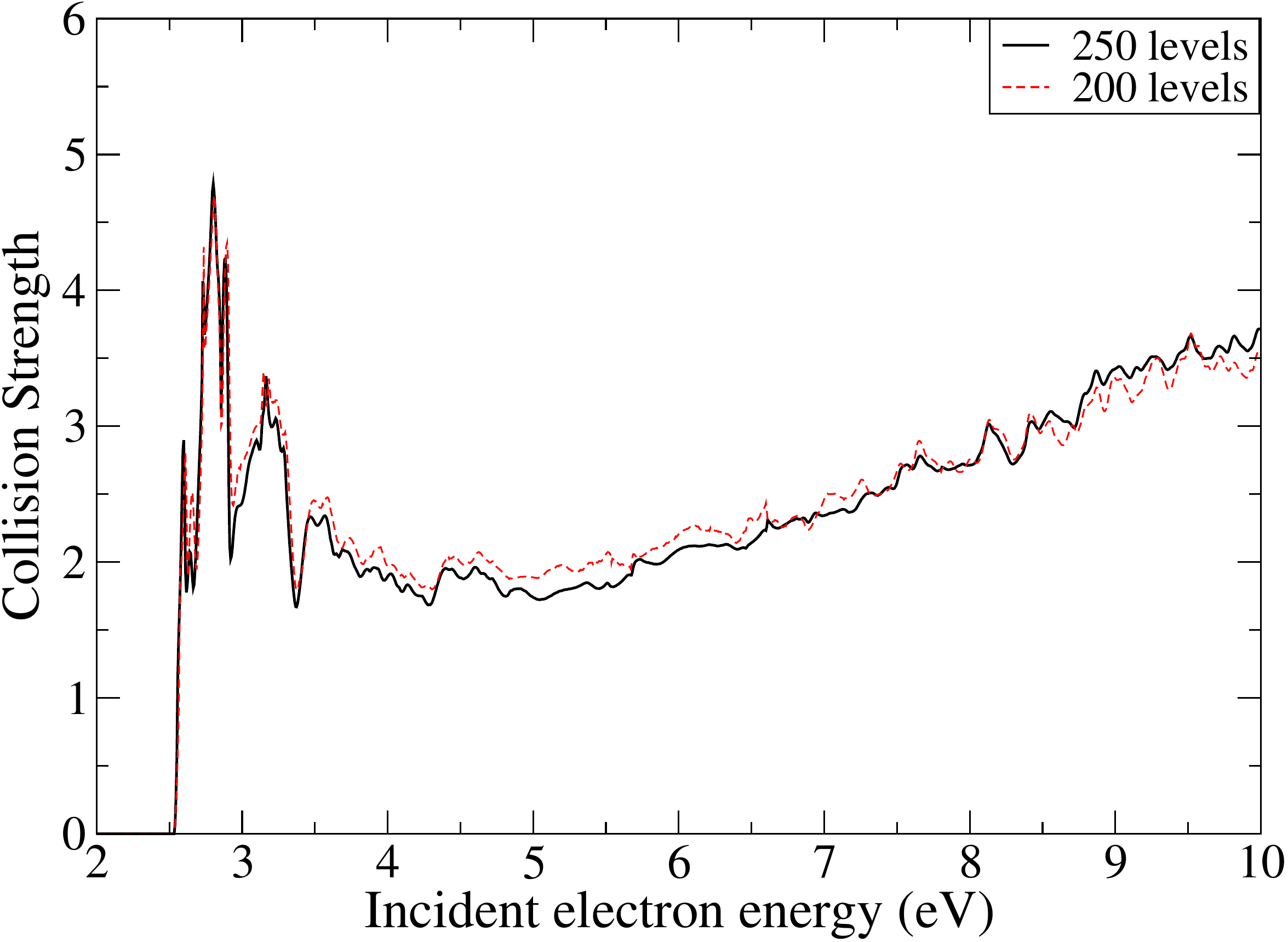}     
\caption{Plot showing collision strengths for the $\ce{{5d}^4{6s}{6p}} \ce{ ^7F^o_5} \rightarrow \ce{{5d}^4{6s}^2} \ce{ ^5D_4}$ transition from the present 250 state R-matrix calculation and the sample 200 state R-matrix calculation. The solid black curve is the current 250 state result and the dashed red curve is the sample 200 state result. \label{488convergence}}
\end{figure}

Additionally, in Fig \ref{488convergence} we present the collision strength for the $\ce{{5d}^4{6s}{6p}} \ce{ ^7F^o_5} \rightarrow \ce{{5d}^4{6s}^2} \ce{ ^5D_4}$ transition from a sample 200 state calculation and compare with the corresponding collision strength from the current 250 state calculation. Comparisons between the two calculations show that there is very good agreement in terms of position and height of the near-threshold resonance, and it is clear that there are no significant changes in either shape or magnitude of the overall collision strengths. We can see that including 50 additional levels in the calculation (taking us from the 200 state model to the 250 state model) results in a small overall average difference of 4.9\% between the collision strengths, indicating near-convergence of the close-coupling expansion. Including even more levels is not expected to change the collision strengths by any significant amount, and will result in even smaller differences overall. As a result, the present comparisons give us confidence in the convergence of the close-coupling expansion and accuracy of our current 250 state scattering calculation. 

From both the partial wave breakdown and sample 200 state calculation we can assert that we have achieved near-convergence and good accuracy for both the present modelling purposes (detailed in the next section) and for future applications of the current scattering model. In the next section we take forward the scattering data and incorporate it into collisional-radiative models to produce a synthetic spectrum, allowing us to directly compare with recently observed spectra from the CTH experiment, details of which are given in the next section. The accuracy of the present calculations will then become even clearer.

\section{Tungsten Spectra}\label{pops} 
As mentioned in Section \ref{section-eie}, to illustrate the use of the present data we build a collisional-radiative model \cite{ref4.1}, with excited level populations normalised to the ground state, using the collision strengths from the scattering calculation along with the radiative transition rates from the atomic structure calculation discussed in Section \ref{section-atomicstructure}. A collisional-radiative matrix $C_{jk}$ is formed by balancing the electron-impact excitation and radiative decay rates, allowing the calculation of the photon emissivity coefficients (PECs), in units of number of photons cm$^{-3}$ s$^{-1}$, defined as
\begin{equation}
\mathcal{PEC}^{(\text{exc})}_{1,j \rightarrow i} =-A_{ji} \sum_{k>1} (C'_{jk})^{-1}C_{k1}.
\end{equation}
Here, $C'_{jk}$ is simply a reduced collisional-radiative matrix with the ground state row removed. We note that the effects of metastable states on the level populations has not been considered here and will be the subject of future work incorporating the present calculations into extensive metastable resolved collisional-radiative models. Previous investigations have focused on diagnostics in the visible region, with one extending this to a small number of lines in the UV region \cite{ref1.4}, but disagreements between theory and observation are seen. The aim of this section is to identify and provide more potential diagnostic lines, aided by comparisons between the current theoretical modelling and recent observations from the CTH experiment.  

Tungsten emission measurements were obtained from the CTH plasma experiment at Auburn University \cite{ref1.8}. Stellarnet survey spectrometers sensitive in the 300-400nm and 400-600nm regions observed emission from plasma-tungsten interactions and were focused on the end of a vertically translating tungsten tipped probe, inserted into the edge of the CTH plasma. For measurements around the tungsten tip the temperature of the CTH plasma is expected to be within the 1-10 eV range and the the electron density is expected to be $\sim$10$^{12}$ cm$^{-3}$.

\begin{figure*}[ht!]
\includegraphics[width=0.8\textwidth]{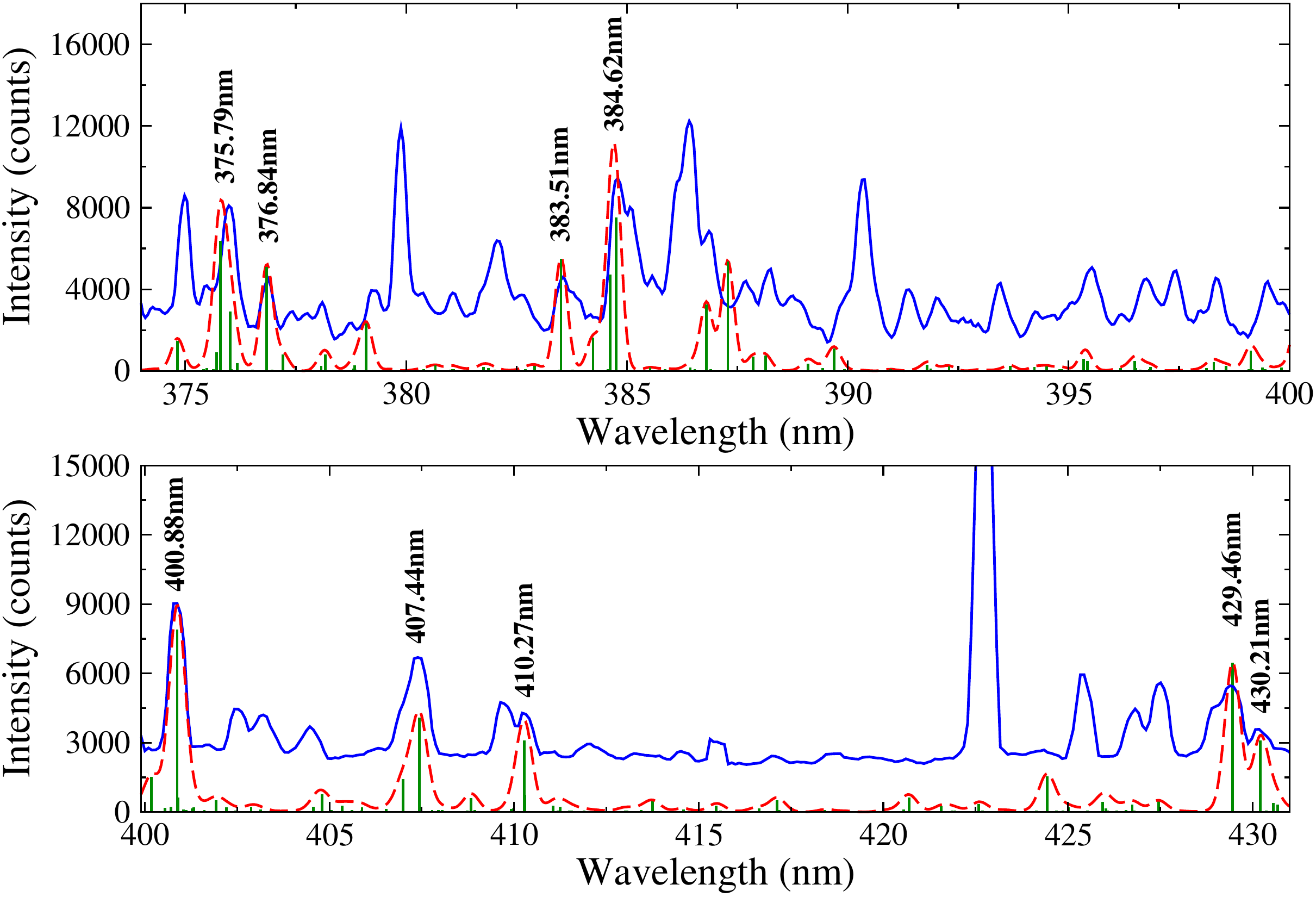}     
\caption{Plot showing the observed spectrum from the CTH experiment (solid blue line) compared to the present theoretical results. Vertical green sticks are the PECs for W I for an electron temperature of 8 eV and electron density of $1 \times 10^{12}$ cm$^{-3}$. The dashed red line is a synthetic spectrum for W I obtained from convolving the PEC data with a Gaussian. \label{spectra}}
\end{figure*}

In Fig \ref{spectra} we provide a sample of the results from the current model compared to observations from the CTH experiment across a 350-450nm range. We see that the well known 400.88nm line, and also the 410.21nm and 430.21nm lines, all exhibit excellent agreement, in terms of both spectral height and position, when comparing the current model with CTH measurements. In addition, reasonable agreement is seen with the 407.44nm line, also previously observed in experiments of Geiger \etal , \cite{ref1.6}, and with the 429.46nm line. It can be seen that the heights of these two lines are not matched exactly with observation. However, this is consistent with the results of the atomic structure calculation, as discussed in Section \ref{section-atomicstructure}, with the calculated radiative transition rate of the 407.44 nm transition being slightly too small, $5.66 \times 10^6$ s$^{-1}$ compared to the experimental value of $1.00\times 10^7$ s$^{-1}$. Similarly for the 429.46 nm transition the radiative transition rate is slightly too large, $3.17\times 10^7$ s$^{-1}$ compared to the experimental value of $1.24 \times 10^7$ s$^{-1}$. Again, referring to Fig \ref{spectra} we see good agreement with observations from the CTH experiment in terms of spectral position for the strong transitions at 375.79nm, 367.84nm, 383.51nm, and at 384.62nm. However, we see that there are slight discrepancies in terms of spectral heights. As before, this is consistent with the results of the atomic structure, which shows that the calculated transition rates are larger than observed values, thus leading to slightly inflated spectral heights than those observed from the CTH experiment. 

As presented in Section \ref{section-atomicstructure}, all lines mentioned above are single transitions from the \ce{^7S_3} and \ce{^5D_J} $(J=0,1,2,3,4)$ metastable levels in the ground state complex to the upper level. It is evident that there is good overall agreement in the wavelength ranges shown, and given the accuracy of the aforementioned atomic structure model, scattering calculations, and collisional-radiative modelling we suggest the above lines as diagnostics and recommend that further experimental investigations be carried out to validate their suitability. A full comparison of the CTH measurements with the present calculations will be the subject of future work. The complete set of data used here will be made available on the OPEN-ADAS website \cite{ref1.21} or at request to the authors.

\section{Conclusions}\label{section-conclusions}
In this work we have investigated the atomic structure and electron-impact excitation of neutral tungsten. The issue of level classification in the literature has been addressed and new non-perturbative calculations of the electron-impact excitation have replaced existing incomplete plane-wave Born and semi-empirical calculations currently used for diagnostic work. Spectral lines across the 350-450nm range have been identified and suggested as diagnostic lines, and they will prove useful for future calculations of tungsten erosion rates.

The atomic structure, determined using the multiconfigurational Dirac-Fock method, was found to be in very good agreement with experiment and corrections to existing level classifications in the literature have been presented with the aid of recent {\it ab initio} calculations. The atomic structure was then carried through to a 250 state Dirac R-matrix calculation and effective collision strengths for a select few transitions are presented. Although these effective collision strengths are compared with the results of a plane-wave Born calculation showing large differences, deducing the accuracy of the R-matrix calculation was difficult given the distinct lack of available non-perturbative data in the literature. However, a breakdown of the partial wave expansion and comparisons with a sample 200 state calculation gave confidence in the convergence and accuracy of the present 250 state calculation. 

Extensive collisional-radiative modelling was employed, incorporating results from both the atomic structure and collisional models, to produce synthetic spectra for neutral tungsten. These allowed direct comparisons of spectral heights and positions with recent measurements from the CTH experiment. Good agreement was seen for strong transitions across the 350-450nm range of the spectrum and alternative diagnostic lines to the widely used 400.88nm line (375.79nm, 367.84nm, 383.51nm, 384.62nm, 407.44nm, 410.21nm, 429.46nm, and 430.21nm) have been suggested. The good agreement seen with new CTH measurements gives credence to the accuracy of the current atomic structure model and to the current large scale Dirac R-matrix scattering calculation for the electron-impact excitation. In addition, both the present calculations and experimental measurements suggest that metastables strongly populate the excited states, enough to contribute to the effective ionisation rates, and therefore the SXB values.

This work presented here will be beneficial for those requiring atomic data of high accuracy for future applications of neutral tungsten. New calculations for the electron-impact ionisation of neutral tungsten, paired with the current electron-impact excitation calculations, will be the subject of future work to determine new SXB spectral diagnostics. These new spectral diagnostics will ultimately help characterise the influx of W I impurities into magnetically confined fusion plasmas.

\section{Acknowledgements} 
R. T. Smyth is funded by the STFC ST/P000312/1 Consolidated Grant and supported by the Professor James Caldwell Travel Scholarship. This work was also supported by the U.S. Department of Energy, Office of Fusion Energy Sciences, under awards DE-SC0015877 and DE-FG02-00ER54610. All calculations were carried out at the Cray XC40 ``Hazelhen'' supercomputer in HLRS Stuttgart and a local cluster at Queen's University Belfast.

%\bibliography{w_paper_refs}
\input{wpaper.bbl}

\end{document}